\documentclass[11pt,letterpaper]{article}
\usepackage{soul}
\usepackage{jheppub}
\usepackage[mathletters]{ucs}
\usepackage[utf8]{inputenc}
\usepackage[T1]{fontenc}
\usepackage{amsfonts}
\usepackage{amsmath}
\usepackage{graphicx,subfigure}
\usepackage{tensor} %%
\usepackage{mathtools} % 
\allowdisplaybreaks[4]
\usepackage{amssymb}
\usepackage{amsthm,amsmath,amssymb}
\usepackage{mathrsfs}
\usepackage{euscript}
\usepackage{color}
\usepackage{braket}
\usepackage{orcidlink}

\newcommand{\rh}{r_{\text{h}}}
\newcommand{\dd}{\text{d}}

\newcommand{\tc}{{\text{c}}}
\title{ \boldmath Topological perspective on bulk boundary thermodynamic equivalence}

\author[a,b]{Si-Jiang Yang\orcidlink{0000-0002-8179-9365},}
\emailAdd{yangsj@lzu.edu.cn}

\author[a,b]{Shan-Ping Wu\orcidlink{0000-0001-6633-1054},}
\emailAdd{120220908841@lzu.edu.cn}

\author[a,b]{Shao-Wen Wei\orcidlink{0000-0003-0731-0610},}
\emailAdd{weishw@lzu.edu.cn}

\author[a,b]{and Yu-Xiao Liu\orcidlink{0000-0002-4117-4176}\footnote{Corresponding author}}
\emailAdd{liuyx@lzu.edu.cn}

\affiliation[a]{Lanzhou Center for Theoretical Physics, Key Laboratory of Theoretical Physics of Gansu Province, Key Laboratory of Quantum Theory and Applications of MoE, Gansu Provincial Research Center for Basic Disciplines of Quantum Physics, Lanzhou University, Lanzhou 730000, China\vspace{0.1cm}}

\affiliation[b]{Institute of Theoretical Physics $\&$ Research Center of Gravitation, School of Physical Science and Technology, Lanzhou University, Lanzhou 730000, China \vspace{0.1cm}}

\abstract{We establish an exact duality between the extended thermodynamics of five-dimensional charged Gauss–Bonnet AdS black holes and the thermodynamic framework of the dual boundary conformal field theory (CFT). The thermodynamics of the dual CFT involves two central charges originating from the trace anomaly. We demonstrate a precise correspondence between the extended first laws on the bulk and boundary sides. Moreover, the topological charges of the CFT thermodynamics, associated with the phase transition and critical point, coincide with those of the corresponding bulk black hole.}

\begin{document}
\maketitle

\flushbottom
%=========================================

%==========================================
\section{Introduction}\label{sec:intro}
%==========================================

The thermodynamics of black holes has remained one of the most significant and active areas of theoretical development in black hole physics. Black holes not only obey the four laws of thermodynamics similar to ordinary thermodynamic systems~\cite{Bardeen:1973gs,Bekenstein:1972tm}, but also they have entropy~\cite{Bekenstein:1973ur} and may possess an underlying microscopic structure~\cite{Strominger:1996sh,Maldacena:1996gb,Wei:2015iwa}, suggesting a deep connection between gravity, quantum theory, and statistical mechanics.

The thermodynamics of black holes in AdS spacetime has been extensively explored, motivated by their rich phase structure and profound connection with the Anti–de Sitter/conformal field theory (AdS/CFT) correspondence. The Hawking–Page phase transition, describing the transition between a Schwarzschild–AdS black hole and a thermal radiation background~\cite{Hawking:1982dh}, can be interpreted as the confinement–deconfinement transition of the quark–gluon plasma in the large-$N$ strongly coupled gauge theory residing on the asymptotic boundary of the bulk spacetime~\cite{Witten:1998zw}. The phase structure of charged AdS black holes exhibits a remarkable resemblance to that of the van der Waals fluid~\cite{Chamblin:1999tk,Chamblin:1999hg}. When the negative cosmological constant is interpreted as the thermodynamic pressure~\cite{Kastor:2009wy}, the resulting small–large black hole phase transition mirrors the liquid–gas transition of the van der Waals system and is characterized by the same set of critical exponents~\cite{Kubiznak:2012wp}. 

Subsequent studies of various AdS black holes within the framework of extended phase space thermodynamics have revealed that their phase behaviors closely parallel those of ordinary thermodynamic systems, featuring phenomena such as van der Waals-like phase transitions~\cite{Zou:2013owa,Liu:2014gvf,Cheng:2016bpx,Xie:2025auj}, reentrant phase transitions~\cite{Frassino:2014pha,Altamirano:2013ane}, triple points~\cite{Altamirano:2013uqa,Wei:2014hba}, isolated critical points~\cite{Ahmed:2022kyv,Hu:2024ldp,Yang:2025xck}, and even superfluid-like behaviors~\cite{Hennigar:2016xwd,Bai:2023woh}. In this formulation of extended black hole thermodynamics, the negative cosmological constant plays the role of thermodynamic pressure, while the black hole mass is naturally identified with the enthalpy of the system rather than its internal energy~\cite{Kubiznak:2016qmn}. Recent progress has further deepened our understanding of the foundations of extended phase space thermodynamics. Notable advances include the development of the extended Iyer–Wald formalism~\cite{Xiao:2023lap,Xiao:2025gle}, an interpretation based on variations of gauge coupling constants~\cite{Hajian:2023bhq}, and the proposal of a higher-dimensional mechanism for realizing a dynamical cosmological constant~\cite{Frassino:2022zaz}. For a different perspective on the interpretation of the negative cosmological constant, see Ref.~\cite{Mancilla:2024spp}.

From a holographic perspective, the role of the negative cosmological constant $\Lambda$ in the extended phase space thermodynamics remains somewhat elusive, as its precise interpretation within the dual CFT has not yet been fully elucidated~\cite{Visser:2021eqk}. Within the framework of the gauge/gravity duality, the thermodynamic behavior of black holes in the bulk spacetime is holographically equivalent to that of a large-$N$, strongly coupled gauge theory residing on the asymptotic boundary of the bulk~\cite{Maldacena:1997re,Witten:1998qj}. However, the interpretation of $\Lambda$ as a thermodynamic pressure in the Smarr relation does not straightforwardly carry over to the boundary theory, since the bulk pressure is not directly dual to the physical pressure in the CFT.

It has been suggested that varying the cosmological constant $\Lambda$ corresponds to varying the number of colors $N$, or equivalently the number of degrees of freedom $N^2$, in the boundary field theory. In conformal field theories, the effective number of degrees of freedom is characterized by the central charge, whose variation drives the system from one CFT to another. Furthermore, since the central charge is related to the AdS radius, a change in the cosmological constant $\Lambda$ naturally induces a variation in the boundary volume. By adopting an appropriately modified boundary metric, a precise boundary formulation of the dual field theory for extended black hole thermodynamics has been established—first for black holes in Einstein gravity~\cite{Ahmed:2023snm,Gong:2023ywu,Ahmed:2023dnh,Baruah:2024yzw}, and later generalized to Lifshitz and hyperscaling-violating black holes within the framework of the Einstein–Maxwell–Dilaton theory~\cite{Cong:2024pvs}. This formulation has even been extended to scenarios with a finite radial cutoff in the bulk and a $T^2$-deformed CFT on the boundary~\cite{Zhang:2025dgm}.

A central charge emerges when the symmetry algebra of a conformal field theory acquires a Weyl anomaly. The number of independent central charges is dictated by the distinct conformal anomalies allowed in the theory, which in turn are determined by the dimensionality of the underlying manifold on which the CFT is defined. In the case of pure Einstein gravity, these central charges are identical~\cite{Henningson:1998gx}. However, once higher-curvature corrections—such as those appearing in quasi-topological or Lovelock gravity—are incorporated, the central charges split and become distinct, encoding richer holographic information. In five-dimensional Einstein gravity, the dual four-dimensional CFT possesses two identical central charges, whereas in more general higher-curvature theories these quantities can differ. Nevertheless, the construction proposed in Ref.~\cite{Ahmed:2023snm} for $D$-dimensional Einstein gravity yields only a single central charge, highlighting its limitation and the potential obstacles in generalizing it to higher-curvature gravitational theories.

While previous studies~\cite{Kumar:2022afq,Afshar:2025sav}, including our own~\cite{Yang:2024krx}, have explored the holographic thermodynamics of Gauss-Bonnet black holes using a single central charge, such approaches rely on a definition ($C \propto L^{d-2}/G_\text{N}$) strictly valid only in Einstein gravity. Because the inclusion of the Gauss-Bonnet term dynamically splits the trace anomaly in the dual four-dimensional CFT, a theoretically rigorous holographic first law must accommodate two distinct central charges.
In this work, we extend the holographic thermodynamic framework to five-dimensional charged black holes in Gauss–Bonnet gravity and formulate the corresponding holographic first law of thermodynamics, which naturally involves two independent central charges. We then analyze the associated thermodynamic behavior and thermodynamical topology, and compare the resulting phase properties and thermodynamic topology with those of the bulk black hole thermodynamics.

The outline of the paper is as follows. In Sec.~\ref{Sec:two}, we formulate the holographic first law of thermodynamics for the dual CFT and analyze its phase structure and critical behavior. In Sec.~\ref{sec:three}, we explore the thermodynamic topology of the dual CFT and compare it with that of the bulk black hole thermodynamics. The last section is devoted to discussion and conclusion.

%===================================
\section{Holographic thermodynamics for the five-dimensional charged Gauss-Bonnet AdS black hole}\label{Sec:two}
%===================================

The thermodynamics for the five-dimensional charged Gauss-Bonnet AdS black hole has been intensively investigated and exhibits a van der Waals-like phase structure~\cite{Wei:2012ui,Cai:2013qga,Lyu:2023sih}.
In this section, we establish an exact correspondence between the thermodynamic properties of the five-dimensional charged Gauss-Bonnet AdS black hole and those of the dual four-dimensional boundary CFT from the bulk thermodynamics, and explore the thermodynamics of the boundary CFT.

%==========================================
\subsection{Extended bulk thermodynamics}
%==========================================

The five-dimensional charged Gauss–Bonnet AdS black hole is an exact solution to the field equations of five-dimensional Gauss–Bonnet gravity with a negative cosmological constant in the presence of an electromagnetic field. The corresponding bulk action is given by~\cite{Boulware:1985wk}
\begin{equation}\label{Action}
  S=\frac{1}{16\pi }\int \dd^5x\sqrt{-g}\left(\mathcal{R}-2\Lambda+\alpha_{\text{GB}}\mathcal{L_{\text{GB}}} +\mathcal{L_{\text{matter}}}\right),
\end{equation}
where the Gauss–Bonnet term takes the form
\begin{equation}\label{GBterm}
\mathcal{L_{\text{GB}}}=\mathcal{R}_{\mu\nu\alpha\beta}\mathcal{R}^{\mu\nu\alpha\beta} -4\mathcal{R}_{\mu\nu}\mathcal{R}^{\mu\nu}+\mathcal{R}^2,
\end{equation}
and the Lagrangian for the matter field is
\begin{equation}
    \mathcal{L_{\text{matter}}}=-4\pi F^{\mu\nu}F_{\mu\nu}.
\end{equation}
Here, $\alpha_{\text{GB}}$ denotes the Gauss–Bonnet coupling constant with the dimension of $[\text{length}]^2$, and the cosmological constant is
\begin{equation}
    \Lambda=-\frac{6}{L^2},
\end{equation}
where $L$ denotes the AdS radius. The electromagnetic field tensor is defined as $F_{\mu\nu}=\partial_{\mu}A_{\nu}-\partial_{\nu}A_{\mu}$, where $A_{\mu}$ denotes the vector potential. 

Within the framework of the low-energy effective action of heterotic string theory, the Gauss–Bonnet term represents the leading quadratic curvature correction to the Einstein action and can be interpreted as a quantum correction to gravity. The Gauss–Bonnet coupling constant $\alpha_{\text{GB}}$ is proportional to the inverse string tension when positive~\cite{Boulware:1985wk,Yang:2020czk}. Therefore, in this work, we restrict our attention to the case of a positive Gauss–Bonnet coupling constant.

The metric for the five-dimensional static spherically symmetric charged Gauss-Bonnet AdS black hole is~\cite{Cai:2001dz,Wiltshire:1985us,Cvetic:2001bk}
\begin{equation}
      \dd s^2=-f(r)\dd t^2+f^{-1}(r)\dd r^2+r^2\dd\Omega_3^2\label{GBmetric}
\end{equation}
with the metric function
\begin{equation}
    \begin{split}
        f(r)=1+\frac{r^2}{2\alpha}\left(1-\sqrt{1+\frac{32\alpha M}{3\pi r^4}-\frac{\alpha Q^2}{3r^6}+\frac{2\alpha \Lambda}{3}}\right),
    \end{split}
\end{equation}
where we have rescaled the Gauss-Bonnet coupling constant by $\alpha=2\alpha_{\text{GB}}$.
The metric~\eqref{GBmetric} describes a five-dimensional static spherically symmetric charged black hole with mass $M$ and charge $Q$. The event horizon $\rh$ of the black hole is determined by the largest positive solution to the equation $f(\rh)=0$. 

In the framework of extended phase space thermodynamics, the negative cosmological constant is regarded as the thermodynamic pressure, and the mass of the black hole is interpreted as enthalpy rather than internal energy~\cite{Kastor:2009wy}. The thermodynamic pressure of the black hole is
\begin{equation}
    P=-\frac{\Lambda}{8\pi}.
\end{equation}
In terms of the radius of the horizon $\rh$, the mass of the black hole can be expressed as
\begin{equation}
    M=\frac{16 \pi ^2 P \rh^6+\pi  Q^2+12 \pi  \alpha  \rh^2+12 \pi  \rh^4}{32 \rh^2}.
\end{equation}
The temperature of the black hole is
\begin{equation}
    T_{\text{H}}=\frac{f'(\rh)}{4\pi}=\frac{32 \pi  P \rh^6-Q^2+12 \rh^4}{48 \pi  \alpha  \rh^3+24 \pi  \rh^5}.
\end{equation}
The entropy of the black hole can be derived from the Euclidean action~\cite{Li:2023men}. It is
\begin{equation}
    S=\frac{3\pi ^2}{2}  \left(\frac{\rh^3}{3}+2 \alpha  \rh\right).
\end{equation}
Due to the presence of the quadratic curvature correction to Einstein gravity, the entropy is different from the usual Benkenstein-Hawking area law.

The electromagnetic potential of the black hole is
\begin{equation}
    \Phi=\left(\frac{\partial M}{\partial Q}\right)_{S,P,\alpha}=\frac{\pi  Q}{16 \rh^2}.
\end{equation}
The thermodynamic volume, conjugate to the thermodynamic pressure, is given by
\begin{equation}
    V=\frac{\pi ^2 \rh^4}{2}.
\end{equation}
To ensure a consistent formulation of both the first law of black hole thermodynamics and the associated Smarr relation for the charged Gauss-Bonnet AdS black hole, it is necessary to regard the Gauss-Bonnet coupling constant $\alpha$ as a thermodynamic variable, whose conjugate quantity can then be derived accordingly:
\begin{equation}
    \mathcal{A}=\left(\frac{\partial M}{\partial \alpha}\right)_{S,P,Q}=\frac{\pi  \left(-32 \pi  P \rh^6+Q^2+6 \alpha  \rh^2-9 \rh^4\right)}{8 \rh^2 \left(2 \alpha +\rh^2\right)}.
\end{equation}
In the extended phase space, a consistent first law of black hole thermodynamics and the Smarr relation for the five-dimensional Gauss-Bonnet AdS black hole are~\cite{Cai:2013qga}
\begin{eqnarray}
  dM &=& TdS+VdP+\Phi dQ+ \mathcal{A}d\alpha, \label{bulk1stlow}\\
  2M &=& 3TS-2PV+2\Phi Q+2\mathcal{A} \alpha.\label{SmarrR}
\end{eqnarray}

Based on the bulk thermodynamics, we derive the boundary CFT thermodynamics in the next subsection.

%==========================================
\subsection{Extended boundary thermodynamics}\label{boundarythermo}
%==========================================

The boundary metric on which the dual CFT resides is given by~\cite{Yang:2024krx,Karch:2015rpa}:
\begin{equation}
	\dd s^2=\omega^2\left(-\dd t^2+L^2\dd\Omega_3^2 \right),
\end{equation}
where $\omega$ is a dimensionless conformal factor. In line with the recent proposal~\cite{Ahmed:2023snm,Ahmed:2023dnh}, rather than fixing $\omega=1$, we adopt the choice $\omega=R/L$, where $R$ denotes the curvature radius of the boundary and is treated as a free parameter. This freedom reflects the underlying conformal symmetry of the boundary theory. Consequently, the spatial volume of the boundary is given by
\begin{equation}
    \mathcal{V}=2\pi^2 R^3.
\end{equation}

In the holographic conformal field theories dual to five-dimensional Gauss–Bonnet gravity, two distinct central charges, 
$C$ and $A$, emerge as trace anomalies when the CFT is placed on a curved background geometry~\cite{Duff:1977ay}. They are explicitly given by~\cite{Myers:2010jv,Myers:2010tj}
\begin{align}
C&=\frac{\pi}{8}\left(1-2 \lambda f_{\infty}\right)  L_{\text{eff}}^3,\label{ccharge} \\
A&=\frac{\pi}{8}\left(1-6 \lambda f_{\infty}\right)  L_{\text{eff}}^3;\label{acharge}
\end{align}
where
\begin{equation}
    \lambda =\frac{\alpha }{L^2}, \quad \quad  f_{\infty}=\frac{1-\sqrt{1-4 \lambda }}{2\lambda},  \qquad L_{\text{eff}}=\sqrt{\frac{2 \lambda }{1-\sqrt{1-4 \lambda }}} L.
\end{equation}
Note that the general formulas derived in Ref.~\cite{Myers:2010jv,Myers:2010tj} also include terms proportional to a curvature-cubed coupling constant, $\mu$. Because we are restricting our analysis strictly to Gauss-Bonnet gravity, which only contains curvature-squared corrections, we have explicitly set $\mu = 0$, leading to the expressions above.

It is important to emphasize that while $C$ and $A$ can be treated as independent thermodynamic variables by simultaneously varying the independent bulk parameters $L$ and $\alpha$, such variations must strictly respect the physical bound $\lambda = \alpha/L^2 \le 1/4$. Variations that violate this constraint fall outside the valid parameter domain and do not correspond to physically permissible CFTs.

Taking into consideration the existence of two central charges, we seek to construct a gauge/gravity correspondence wherein the first law of thermodynamics on the field theory side is represented as
\begin{equation}
d\Tilde{E}=\Tilde{T}d\Tilde{S}+\Tilde{\phi}d\Tilde{Q}+\mu_C dC+\mu_{A}dA-\mathcal{P}d\mathcal{V},
\end{equation}
where $\Tilde{E}$ denotes the internal energy of the boundary field theory; $\Tilde{T}$ and $\Tilde{S}$ are its temperature and entropy; $\Tilde{\Phi}$ and $\Tilde{Q}$ represent the electric potential and charge; $\mathcal{P}$ and $\mathcal{V}$ are its pressure and volume, and $\mu_{\text{C}}$ and $\mu_{\text{A}}$ are the chemical potentials conjugate to the central charges $C$ and $A$, respectively. It is worth noting that varying the central charges $C$ and $A$ corresponds to tracing different CFTs within a continuous family rather than describing fluctuations within a single thermodynamic ensemble. Consequently, the conjugate chemical potentials $\mu_C$ and $\mu_A$ do not couple to local boundary operators, but rather represent the thermodynamic cost of altering the underlying UV degrees of freedom of the dual field theory.

The Euler relation, which serves as the dual counterpart of the Smarr relation, takes the form
\begin{equation}
\Tilde{E}=\Tilde{T}\Tilde{S}+\Tilde{\phi}\Tilde{Q}+\mu_C C+\mu_{A}A.
\end{equation}
It is important to note that the pressure and volume terms do not appear in the Euler relation.

The holographic dictionary relating the thermodynamic variables of the boundary CFT to the bulk is
\begin{equation}
\begin{split}
    \Tilde{E}&=\frac{M}{\omega}, \qquad \Tilde{T}=\frac{T}{\omega}, \qquad \Tilde{S}=S, \\
   \Tilde{\Phi}&=\frac{\Phi }{ \omega L}, \qquad \Tilde{Q}=Q L.\label{dictionary}
\end{split}
\end{equation}
The thermodynamic pressure $\mathcal{P}$, and the chemical potentials $\mu_{C}$ and $\mu_{A}$ conjugate to the thermodynamic volume $\mathcal{V}$ and the central charges 
$C$ and $A$, respectively, are defined as
\begin{equation}
  \mathcal{P}=\frac{\Tilde{E}}{3\mathcal{V}}, \qquad   \mu_{\text{C}}=\left(\frac{\partial \Tilde{E}}{\partial C}\right)_{S,\Tilde{Q},\mathcal{V},A}, \qquad \mu_{\text{A}}=\left(\frac{\partial \Tilde{E}}{\partial A}\right)_{S,\Tilde{Q},\mathcal{V},C}.
\end{equation}
Given the complex and lengthy nature of the exact expressions for the chemical potentials $\mu_C$ and $\mu_A$, we have deferred their explicit presentation to the Appendix.

The holographic dictionary~\eqref{ccharge},~\eqref{acharge}, and~\eqref{dictionary}, together with the thermodynamic quantities of the black hole, provide information about the thermodynamics of the dual large-$N$, strongly coupled CFT. We can express any thermodynamic variables of the CFT using the five independent thermodynamic variables $\Tilde{S}, \Tilde{Q}, \mathcal{V}, C$ and $A$. For this purpose, we define the following two dimensionless parameters
\begin{equation}
    \begin{split}
        x=\frac{\rh}{L},\qquad y=\frac{Q}{L^2}.
    \end{split}
\end{equation}
Then, the thermodynamics of the CFT can be expressed as:
\begin{itemize}
 \item internal energy:
    \begin{equation}
            \Tilde{E}=\frac{\pi  L^3 \left(12\lambda x^2+12 x^6+12 x^4+y^2\right)}{32 R x^2},
    \end{equation}
\item temperature and entropy:
\begin{align}
    \Tilde{T}&=\frac{24 x^6+12 x^4-y^2}{48 \pi\lambda R x^3+24 \pi  R x^5},\\
    \Tilde{S}&=\frac{1}{2} \pi ^2 L^3 x \left(6\lambda+x^2\right),
\end{align}
\item electrostatic potential and charge:
\begin{align}
    \Tilde{\Phi}=\frac{\pi  y}{16 R x^2},\qquad \Tilde{Q}=L^3 y,
\end{align}
\item thermodynamic pressure:
\begin{equation}
     \mathcal{P}=\frac{L^3 \left(12 \lambda x^2+12 x^6+12 x^4+y^2\right)}{48 \pi  R^4 x^2}.
\end{equation}
\end{itemize}
With the boundary thermodynamics established, we proceed to investigate its critical point and phase structure.
%==============================

%==============================
\subsection{Critical point and phase transition for the boundary thermodynamics}\label{critphast}
%==============================

The thermodynamics of the five-dimensional charged Gauss–Bonnet AdS black hole features a characteristic swallowtail structure below the critical point, indicative of a first-order phase transition, which disappears when the parameters are larger than their corresponding critical values~\cite{Wei:2014hba}. In this subsection, we examine the thermodynamic phase transitions of the dual CFT on the boundary and compare them with those of the black hole.

\begin{figure*}
    \centering
    \subfigure[]{\includegraphics[width=0.47 \textwidth]{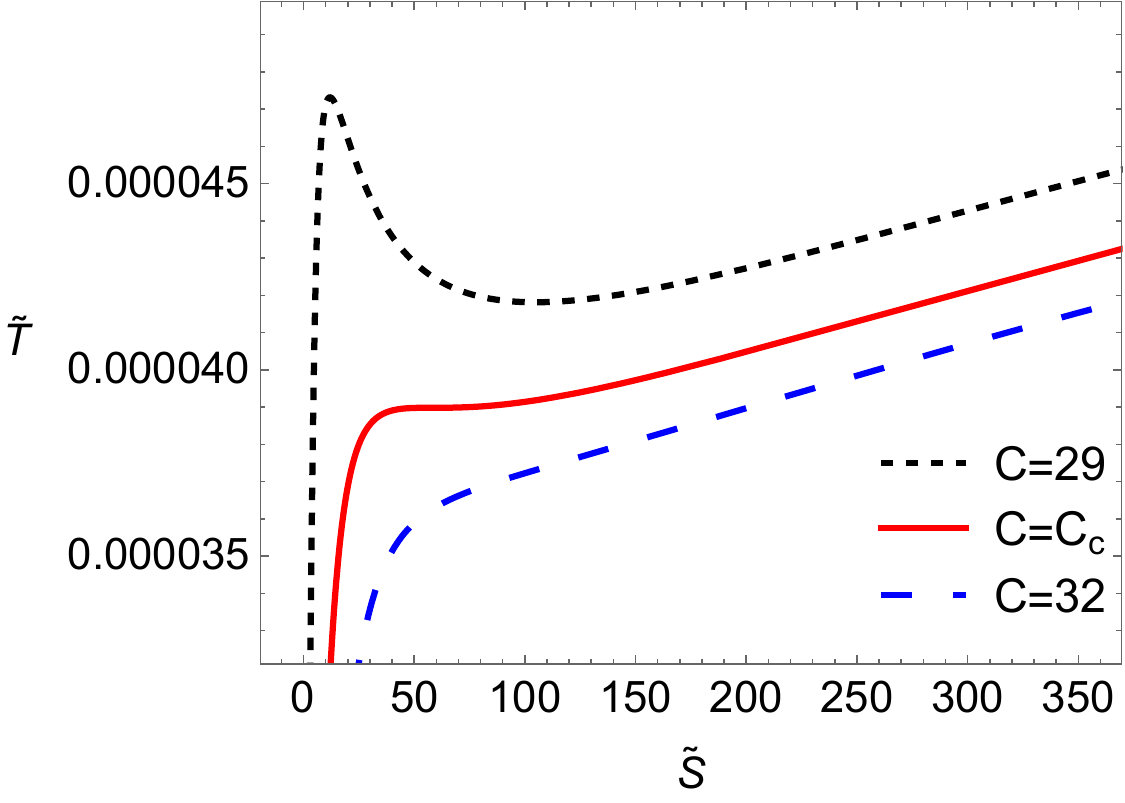}\label{fig:TSN}}
    \subfigure[]{\includegraphics[width=0.47 \textwidth]{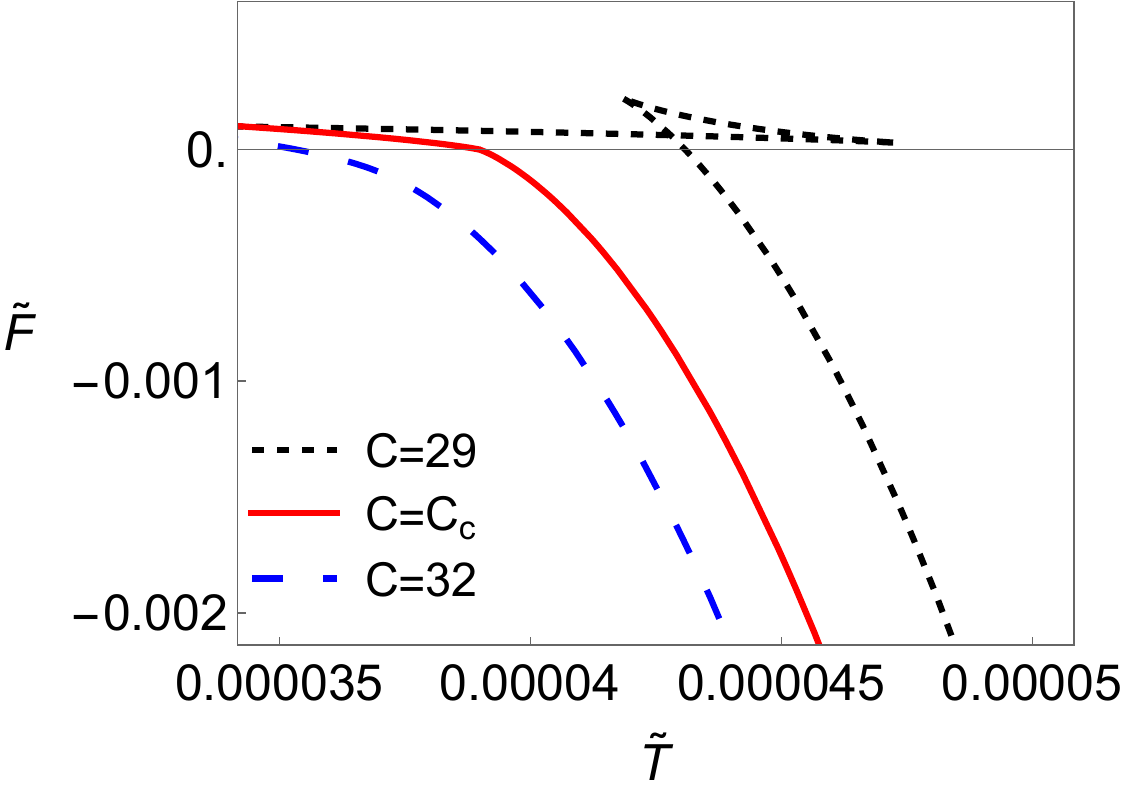}\label{fig:FTN}}
    \caption{The $\Tilde{T}-\Tilde{S}$ oscillatory behavior and the swallowtail behavior of the dual CFT. Here we have set the charge $\Tilde{Q}=0$, the boundary radius $R=10000$, and the central charge $A=27$.}
    \label{Nphase}
\end{figure*}

\begin{figure*}
    \centering
    \subfigure[]{\includegraphics[width=0.47 \textwidth]{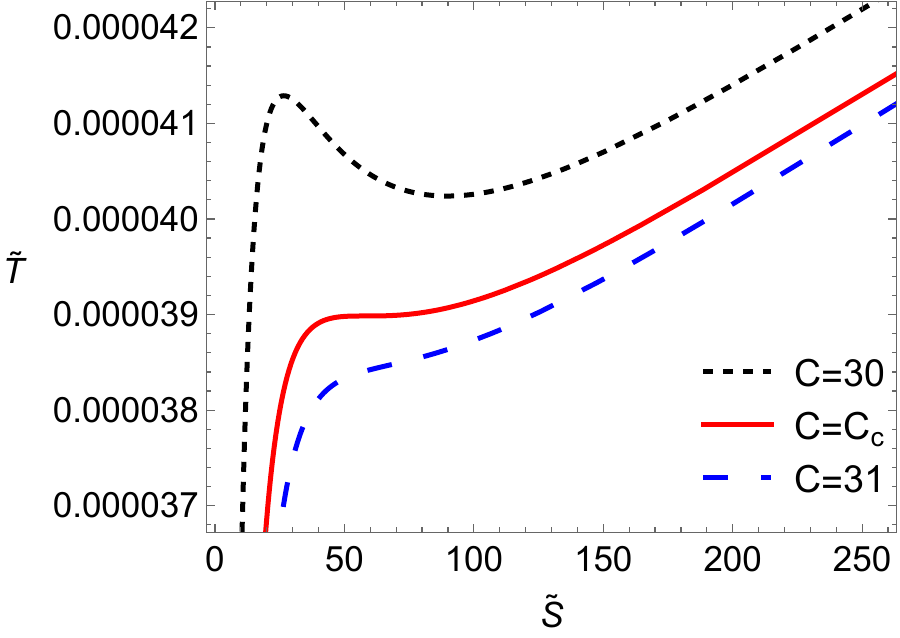}\label{fig:TSC}}
    \subfigure[]{\includegraphics[width=0.47 \textwidth]{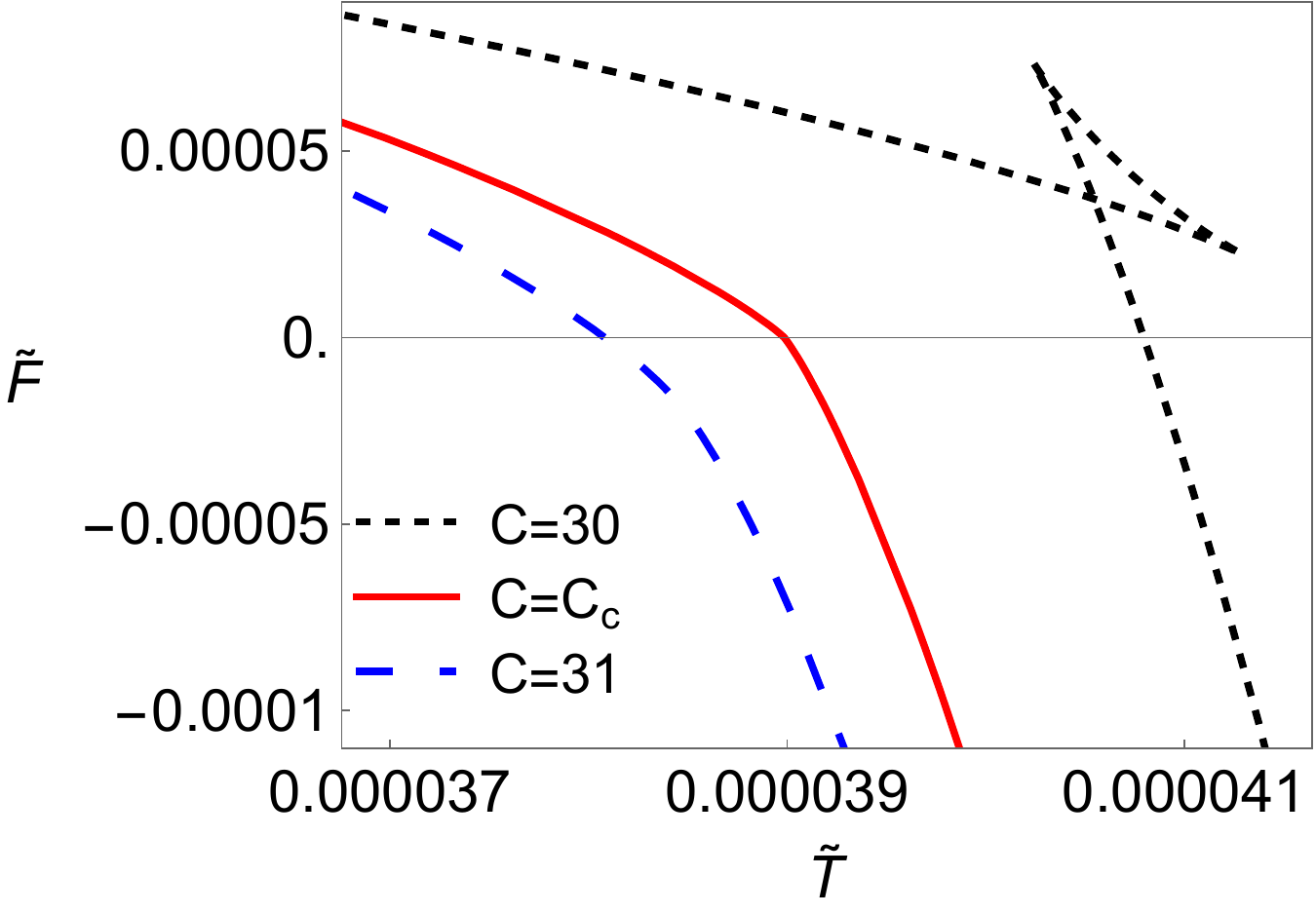}\label{fig:FTC}}
    \caption{The $\Tilde{T}-\Tilde{S}$ oscillatory behavior and the swallowtail behavior of the dual CFT. Here we have set the charge $\Tilde{Q}=0.0866855$, the boundary radius $R=10000$, and the central charge $A=27$.}
    \label{Cphase}
\end{figure*}

Analogous to the extended phase space thermodynamics of the five-dimensional charged Gauss-Bonnet AdS black hole, the dual CFT in the fixed $(\Tilde{Q},C,A,\mathcal{V})$ ensemble exhibits an oscillatory curve in the $\Tilde{T}-\Tilde{S}$ plane and a characteristic swallowtail structure in the free energy below the critical point. Both the oscillatory behavior and the swallowtail structure disappear once the system surpasses the critical point. Figures~\ref{Nphase} and~\ref{Cphase} illustrate the thermodynamic behavior of the dual CFT for $\Tilde{Q}=0$ and $\Tilde{Q}\neq 0$, respectively.

The critical point of the boundary CFT is determined by the conditions
\begin{equation}
    \begin{split}
       \left( \frac{\partial\Tilde{T}}{\partial\Tilde{S}}\right)_{\Tilde{Q},C,A,\mathcal{V}}=0, \qquad   \left(\frac{\partial^2\Tilde{T}}{\partial^2\Tilde{S}}\right)_{\Tilde{Q},C,A,\mathcal{V}}=0.
    \end{split}
\end{equation}
Solving these equations leads to the following expressions for the dimensionless parameters of the boundary CFT at the critical point:
\begin{align}
        x_{\text{c}}&= \frac{1}{6} \left(2^{2/3} K+\frac{2 \sqrt[3]{2}}{K}+2\right)^{\frac{1}{2}},\label{criticala}\\
        \lambda_{\text{c}} &= \frac{2^{2/3}\left[ K^3 \left(28-1215 y^2\right)-32\right]+6 K^2 \left(2835 y^2+4\right)-16 \sqrt[3]{2} K^4+2 \sqrt[3]{2} K \left(28-1215 y^2\right)}{648 K^2 \left(243 y^2+4\right)},\label{criticalb}
\end{align}
where we have defined $K=\sqrt[3]{9 y \left(\sqrt{18225 y^2+60}+135 y\right)+2}$ to simplify the expressions.
%================================
%================================
%================================
%================================

Based on the values of the dimensionless parameters at the critical point, one can determine the corresponding critical thermodynamic quantities for the boundary CFT. For the CFT dual to the five-dimensional neutral Gauss-Bonnet AdS black hole, these critical parameters take the form
\begin{equation}
    \begin{split}
        x_{\text{c}}= \frac{1}{\sqrt{6}}, \qquad \lambda_{\text{c}} =\frac{1}{36}.\label{ncp}
    \end{split}
\end{equation}
The corresponding critical thermodynamic quantities for the boundary CFT are
\begin{equation}
    \begin{split}
        \Tilde{T}_{\tc}=\frac{\sqrt{6}}{2\pi  R}, \qquad C_{\tc}=\frac{1}{3} \left(\sqrt{2}+2\right)A.
    \end{split}
\end{equation}
A straightforward calculation based on Eq.~\eqref{ncp} shows that the critical point of the boundary theory coincides with that of the five-dimensional neutral Gauss–Bonnet AdS black hole~\cite{Cai:2013qga,Yang:2024krx}
\begin{equation}\label{BHcritical}
    r_{\text{hc}}=\sqrt{6\alpha}, \qquad P_c=\frac{1}{48\pi\alpha},  \qquad T_c=\frac{1}{2\sqrt{6\alpha}\pi}.
\end{equation}

For the charged case with $\Tilde{Q}\neq 0$, the corresponding critical point for the boundary CFT is
\begin{align}
    \Tilde{T}_{\text{c}}=\frac{24 x_{\text{c}}^6+12 x_{\text{c}}^4-y^2}{48 \pi\lambda_{\text{c}} R x_{\text{c}}^3+24 \pi  R x_{\text{c}}^5}.
\end{align}
Further calculations confirm consistency with the critical point of the bulk thermodynamics, with the typographical errors in Ref.~\cite{Cai:2013qga} duly corrected.

%==============================
\section{Topology of the CFT thermodynamics}\label{sec:three}
%==============================

Understanding the universal aspects shared by the bulk and the dual boundary CFT thermodynamics remains an issue of enduring interest. Topological methods offer a compelling avenue to reveal such deep connections~\cite{Wei:2021vdx,Wei:2022dzw,Wu:2025xxo,Wei:2024gfz,Wu:2024asq,Fan:2022bsq,Babaei-Aghbolagh:2025qxm}. Recent studies based on the residue method~\cite{Fang:2022rsb} have demonstrated that the charged AdS black hole in Einstein gravity and its dual CFT with a central charge share identical topological charges associated with the first-order phase transition and the critical point~\cite{Zhang:2023uay}. Extending this line of research, we systematically investigate the thermodynamic topology of the boundary CFT dual to the five-dimensional charged AdS black hole, with particular emphasis on the topological characteristics of its phase transitions and critical point. We further contrast the thermodynamic topology of the boundary CFT with those of the corresponding bulk black hole, thereby uncovering their intrinsic topological correspondence.

%==============================
\subsection{Thermodynamic topological charges}
%==============================

Topology provides a systematic and robust framework for classifying thermodynamic behavior in terms of topological invariants. To assign a topological charge to a thermodynamic process, we introduce a unit vector field $n^a=\phi^a/||\phi||$ constructed from a two-dimensional vector field $\phi$ defined in a two-dimensional parameter space. Following Duan’s $\phi$-mapping topological current theorem~\cite{Duan:1979ucg}, the corresponding topological current can be defined as
\begin{equation}
    j^\mu=\frac{1}{2\pi}\epsilon^{\mu\nu\alpha}\epsilon_{ab}\partial_\nu n^a 
    \partial_\alpha n^b, \qquad \mu, \nu, \alpha=0, 1, 2.
\end{equation}
where $\partial_\nu=\partial/\partial x^\nu$ and $x^\nu=(t,\Tilde{S},\Theta)$.
This current is identically conserved,
\begin{equation}
    \partial_\mu j^\mu=0,
\end{equation}
reflecting the topological nature of the construction and ensuring that the associated topological charge remains invariant under smooth deformations of the underlying thermodynamic parameters.

Utilizing the two-dimensional Laplacian Green’s function $\Delta_{\phi} \ln ||\phi||=2\pi \delta^2(\phi)$ and the Jacobian tensor $\epsilon^{ab}J^\mu(\frac{\phi}{x})=\epsilon^{\mu\nu\alpha}\partial_\nu \phi^a \partial_\alpha n^b$, the topological current can be written in a compact form as
\begin{equation}
    j^\mu=\delta^2(\phi) J^\mu\left(\frac{\phi}{x}\right).
\end{equation}
This expression highlights a key feature of the topological current: it is nonvanishing only at the zero points of the vector field $\phi$, where the topology of the field configuration becomes nontrivial.
Denote the $i$-th zero point of the vector field $\phi$ by $\textbf{z}_i $. Then the density of the topological current can be written as
\begin{equation}
    j^0= \sum_i \beta_i \eta_i \delta^2(\textbf{x}-\textbf{z}_i),
\end{equation}
where $\beta_i$ is the Hopf index, and $\eta_i=\text{sign}\left(J^0(\phi/x) \right)$ represents the Brouwer degree of the mapping at the zero point $\textbf{z}_i$.

The conservation of the current implies the existence of a topological charge $\mathcal{Q}$, defined by
\begin{equation}
    \mathcal{Q}=\int_\Sigma j^0 d^2x=\sum_i \beta_i\eta_i=\sum_i w_i,
\end{equation}
where $w_i$ denotes the winding number associated with the $i$-th zero point of the vector field $\phi$.

%==============================
\subsection{Topology of the first-order phase transition}
%==============================

We consider a canonical ensemble of the boundary CFT thermodynamics, in which $(\Tilde{Q},\mathcal{V}, C, A)$ are held fixed. The order parameter characterizing the CFT thermodynamic phase transition is defined as the difference between the entropy of the dual CFT and its value at the critical point. For simplicity, we hereafter refer to this quantity as the CFT entropy.
Define the generalized free energy~\cite{York:1986it,Yang:2021ljn} of the boundary CFT as
\begin{equation}
\begin{split}
    \mathcal{F}&=\Tilde{E}-\frac{\Tilde{S}}{\tau}\\
    &=\frac{\pi  L^3 \left(12 \lambda  x^2+12 x^6+12 x^4+y^2\right)}{32 R x^2}-\frac{\pi ^2 L^3 x \left(6 \lambda +x^2\right)}{2 \tau },\label{offshellF}
\end{split}
\end{equation}
where $\Tilde{E}$ and $\Tilde{S}$ are the energy and entropy of the dual CFT. The parameter $\tau$ is introduced as an auxiliary variable with the dimension of time, corresponding to the inverse ensemble temperature of the boundary CFT. We allow $\tau$ to vary freely; in this case, the generalized free energy is generally off-shell. The on-shell state arises only at the extremum of the generalized free energy in the $\mathcal{F}-\Tilde{S}$ plane, located at $\tau=1/\Tilde{T}$.

To investigate the first-order phase transition of the boundary CFT, we define the vector field $\phi=(\phi^{\Tilde{S}},\phi^{\Theta})$ as~\cite{Wei:2022dzw}
\begin{align}
   \phi^{\Tilde{S}}&= \left(\frac{\partial \mathcal{F}}{\partial \Tilde{S}}\right)_{\Tilde{Q},\mathcal{V},C,A},\\
  \phi^{\Theta}&= -\frac{\cot\Theta\csc\Theta}{R},
\end{align}
where $R$ is the radius of the boundary. The parameter $\Theta$ ranges as $0\leq \Theta\leq \pi$. Unlike some of our authors’ previous work~\cite{Wei:2022dzw}, the present definition of the component $\phi^{\Theta}$ introduces a denominator $R$. This denominator does not affect the winding number of the vector field; rather, it serves to render the variation of the vector field smoother instead of abrupt.

Explicit calculations reveal that the component $\phi^{\Tilde{S}}$ takes the form
\begin{equation}
    \phi^{\Tilde{S}}=\frac{24 x^6+12 x^4-y^2}{24 \pi  R x^5+48 \pi  \lambda  R x^3}-\frac{1}{\tau }.
\end{equation}
At  $\Theta=0$ and  $\Theta= \pi$, the component $\phi^{\Theta}$ diverges, causing the vector field to point outward. The zero point of the vector field $\phi$ is located at
\begin{align}
\left\{\begin{lgathered}
    \tau=\frac{24 \pi  R x^5+48 \pi  \lambda  R x^3}{24 x^6+12 x^4-y^2}=\frac{1}{\Tilde{T}},\\
    \Theta=\frac{\pi}{2}.
\end{lgathered}\right.
\end{align}
This confirms that the free energy of the CFT coincides with the zero point of the vector field $\phi$. From a topological perspective, this allows us to assign a topological charge via the vector field $\phi$. 

\begin{figure*}
    \centering
    \subfigure[]{\includegraphics[width=0.47 \textwidth]{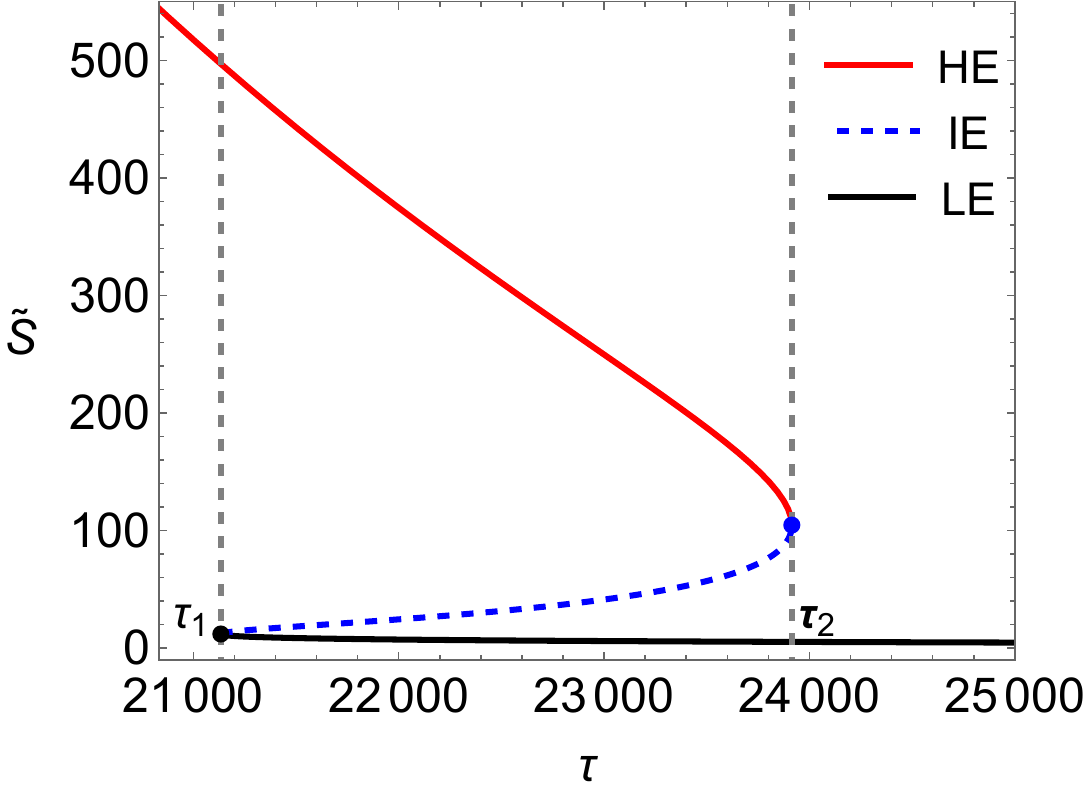}\label{fig:zpneutral}}
    \subfigure[]{\includegraphics[width=0.47 \textwidth]{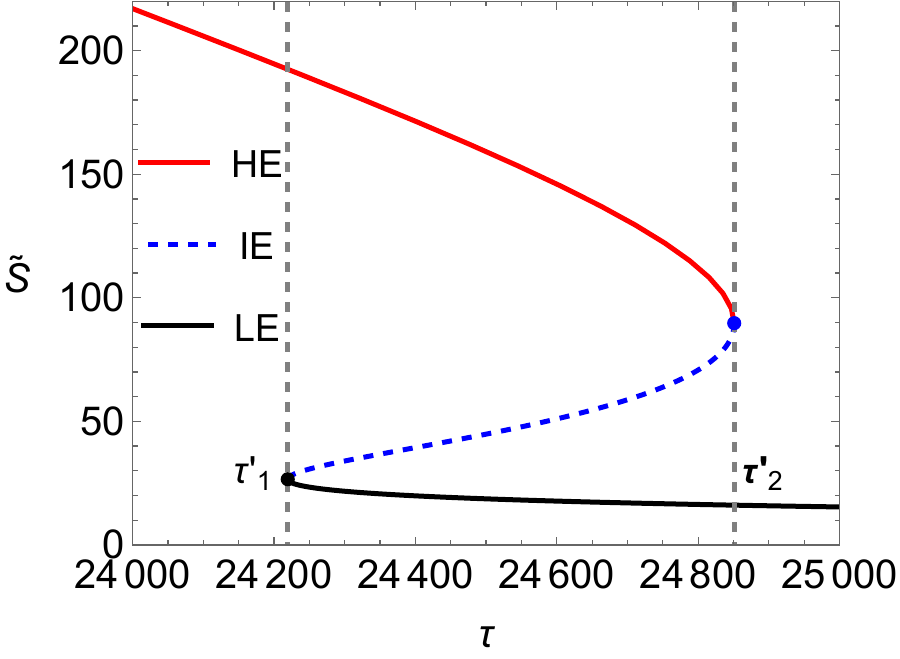}\label{fig:zpcharged}}
    \caption{ Zero points of $\phi^{\tilde{S}}$ in the $\tau$--$\tilde{S}$ plane for the dual CFT. 
        The black solid, blue dashed, and red solid curves correspond to the low-entropy (LE), intermediate-entropy (IE), and high-entropy (HE) branches, respectively. Black and blue dots indicate the annihilation and generation points. We have taken the boundary radius $R=10,000$, the central charge $A=27$. (a) Zero points of $\phi^{\tilde{S}}$ in the $\tau$--$\tilde{S}$ plane for the neutral dual CFT ($\tilde{Q}=0$) with central charge $C=29$.  (b) Zero points of $\phi^{\tilde{S}}$ in the $\tau$--$\tilde{S}$ plane for the charged dual CFT with electric charge $\tilde{Q}=0.0866855$ and central charge $C=30$. }
    \label{zps}
\end{figure*}

Taking the boundary radius $R=10000$ and the central charge $A=27$, we exhibit zero points of $\phi^{\Tilde{S}}$ in the $\Tilde{S}-\tau$ plane for neutral and charged dual CFT in Figs.~\ref{fig:zpneutral} and~\ref{fig:zpcharged}, respectively.
%================================
%================================
As shown in the figures, when the central charge is below its critical central charge, the curve decomposes into three piecewise-smooth branches. The black solid line represents the low-entropy (LE) branch, the blue dashed line corresponds to the intermediate-entropy (IE) branch, and the red solid line denotes the high-entropy (HE) branch. 
There exists only a single dual CFT state with high entropy for $\tau<\tau_1$ or $\tau<\tau'_1$. 
In the interval $\tau_1<\tau<\tau_2$ or $\tau'_1<\tau<\tau'_2$, three distinct entropy states coexist.
For $\tau>\tau_2$ or $\tau>\tau'_2$, only a single low-entropy state remains. 

\begin{figure*}
    \centering
    \subfigure[]{\includegraphics[width=0.47 \textwidth]{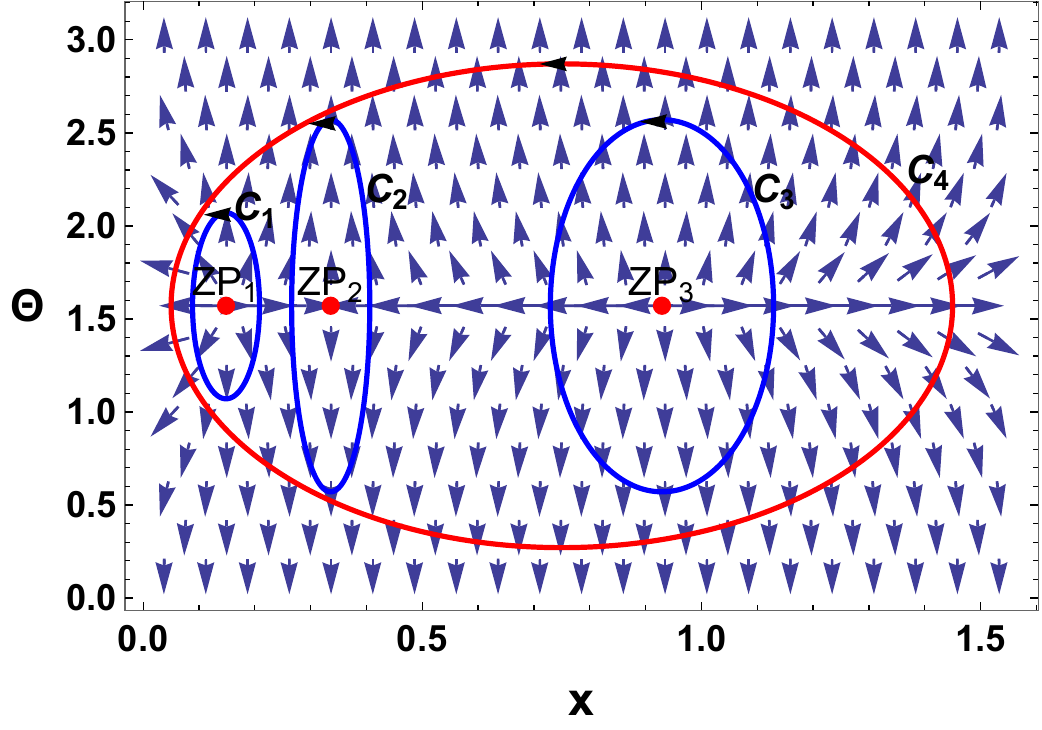}\label{fig:vecfneutral}}
    \subfigure[]{\includegraphics[width=0.47 \textwidth]{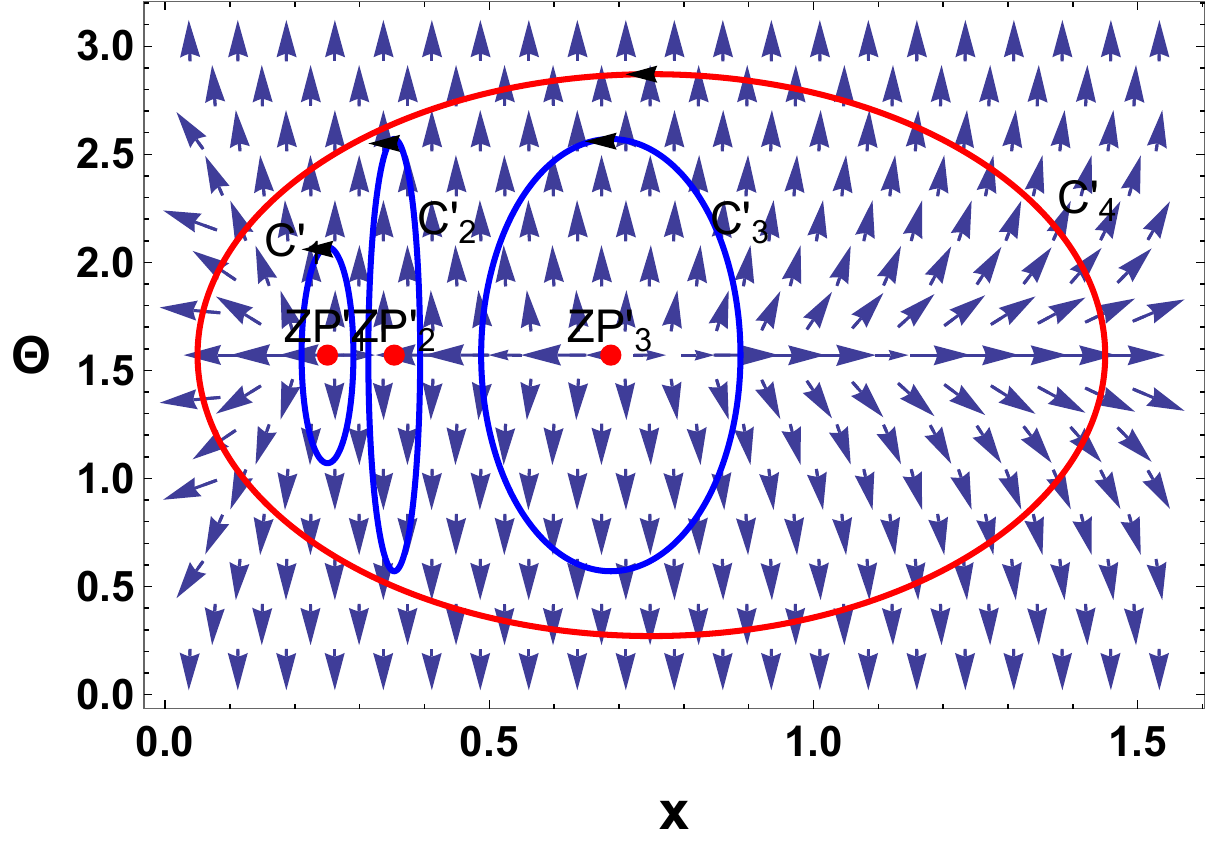}\label{fig:vecfcharged}}
    \caption{The arrows depict the unit vector field $\mathbf{n}$ on a portion of the $x-\Theta$ plane, with zero points indicated by red dots. (a) For the dual CFT with $\Tilde{Q}=0$, the zero points of the vector field are located at $(x,\Theta)=(0.148427,\pi/2)$, $(0.335925,\pi/2)$, and $(0.929365,\pi/2)$, corresponding to ZP$_1$, ZP$_2$, and ZP$_3$, respectively. These points are associated with the low-, intermediate-, and high-entropy states of the dual CFT. (b) For the dual CFT with $\Tilde{Q}\neq 0$, the zero points of the vector field are located at $(x,\Theta)=(0.250441,\pi/2)$, $(0.353396,\pi/2)$, and $(0.687455,\pi/2)$ for the charge value $\Tilde{Q}=0.0866855$, corresponding to ZP$'_1$, ZP$'_2$, and ZP$'_3$, respectively. These points again represent the low-, intermediate-, and high-entropy states of the dual CFT. Throughout the analysis, we set the boundary radius to $R=10000$ and the central charge to $A=27$.}
    \label{phase1st}
\end{figure*}

For $\tau_1<\tau<\tau_2$ or $\tau'_1<\tau<\tau'_2$, we present the unit vector field $n=(\phi^{\Tilde{S}}/||\phi||, \phi^{\Theta}/||\phi||)$  over a representative portion of the $\Theta-x$ plane for neutral and charged dual CFT in Figs.~\ref{fig:vecfneutral} and~\ref{fig:vecfcharged}, respectively, where $|| \phi ||$ serves as the normalization factor.
%================================
%================================
The red points in the figures mark the zero points of the vector field, which correspond to the three branches of the dual CFT. Since the winding number $w$ is independent of the chosen loop enclosing a zero point, it can be evaluated using any convenient contour. 
%================================
\begin{figure*}
    \centering
    \subfigure[]{\includegraphics[width=0.47 \textwidth]{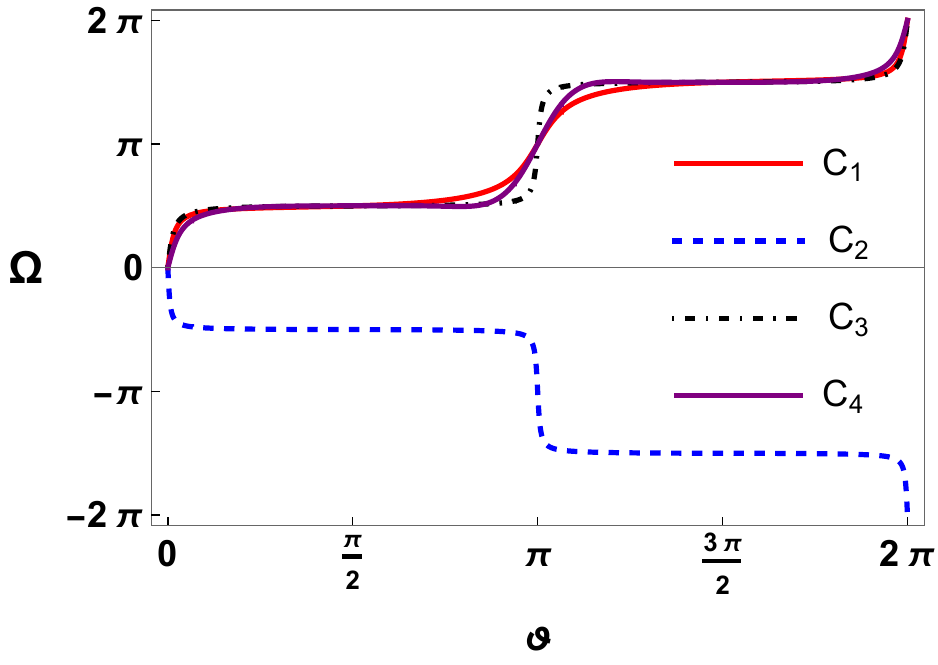}\label{fig:wnneutral}}
    \subfigure[]{\includegraphics[width=0.47 \textwidth]{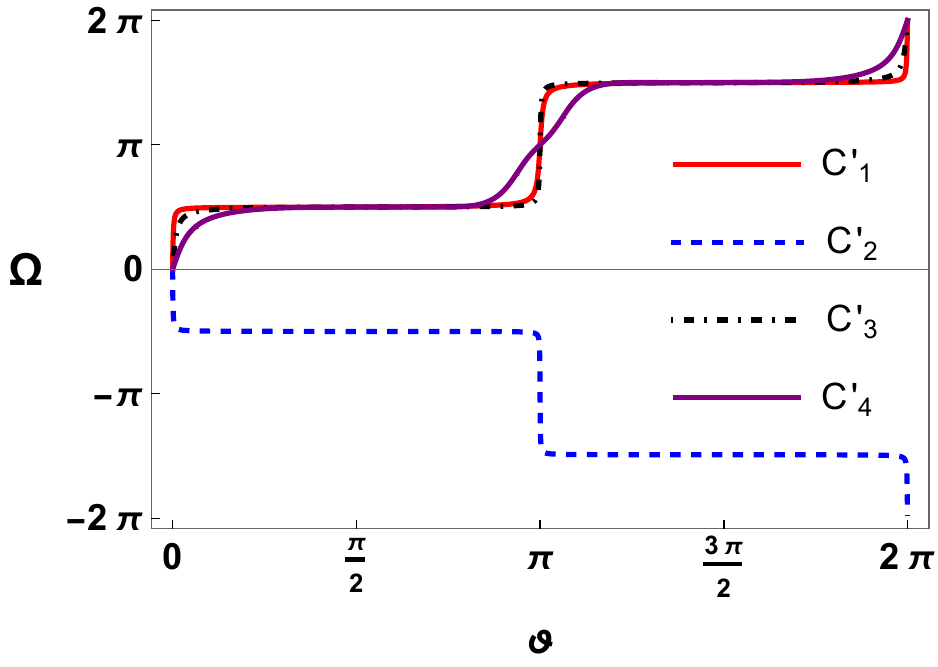}\label{fig:wncharged}}
    \caption{$\Omega$ vs.~$\vartheta$ for the contours $C_{1}, C_{2}, C_{3}, C_{4}$ in subfigure~(a) 
(with $\tilde{Q} = 0$) and $C'_{1}, C'_{2}, C'_{3}, C'_{4}$ in subfigure~(b) 
(with $\tilde{Q} \neq 0$).}
    \label{wn1s}
\end{figure*}
%================================

For convenience in computing the topological charge, we introduce a quantity that measures the deflection of the vector field along the contour, defined as
\begin{equation}
    \Omega(\vartheta)=\int^{\vartheta}_0\epsilon_{ab}n^a\partial_{\vartheta}n^b d\vartheta.
\end{equation}
The corresponding topological charge is then given by
\begin{equation}
    \mathcal{Q}=\frac{1}{2\pi}\Omega(2\pi).
\end{equation}

For the three branches of the CFT states, we find that the low-entropy and high-entropy branches both yield $w=1$, while the intermediate-entropy branch yields $w=-1$. The values of the topological charge obtained from integrals along different contours are shown in Figs.~\ref{fig:wnneutral} and~\ref{fig:wncharged}. Consequently, at this value of the central charge, the dual CFT possesses a total topological charge $\mathcal{Q}=1-1+1=1$, which remains invariant with respect to the choice of $\tau$. Therefore, the topological charge 
$\mathcal{Q}$ is always one for the phase-transition thermodynamics of the boundary CFT dual to the five-dimensional charged Gauss–Bonnet AdS black hole, in agreement with the thermodynamic topological charge of the black hole itself~\cite{Liu:2022aqt,Jeon:2024yey}.

%================================
\subsection{Topology of the critical point}
%================================

At the critical point, the entropy difference between the low-entropy and high-entropy phases disappears, and the two phases become indistinguishable. The physics in the vicinity of this point is of particular importance, as it typically exhibits the universal behavior characteristic of critical phenomena. As demonstrated in subsection~\ref{critphast}, the thermodynamics of the dual CFT admits a critical point within the fixed $(\Tilde{Q},\mathcal{V}, C, A)$ canonical ensemble. In this subsection, we explore the thermodynamic topology associated with this critical point.

The temperature of the boundary CFT can be regarded as a function of the entropy $\Tilde{S}$, the electric charge $\Tilde{Q}$, the central charge $C$, the central charge $A$, and the boundary volume $\mathcal{V}$:
\begin{equation}
    \Tilde{T}=\Tilde{T}(\Tilde{S},\Tilde{Q},C,A,\mathcal{V}).
\end{equation}
Requiring the condition 
\begin{equation}
\left(\frac{\partial\Tilde{T}}{\partial\Tilde{S}}\right)_{\Tilde{Q},C,A,\mathcal{V}}=0,
\end{equation}
we can eliminate the parameter $\lambda$ and express the temperature of the CFT  as
\begin{equation}
    \Tilde{T}=\frac{24 x^6+4 x^4+y^2}{16 \pi  R x^5}.
\end{equation}

\begin{figure*}
    \centering
    \subfigure[]{\includegraphics[width=0.47 \textwidth]{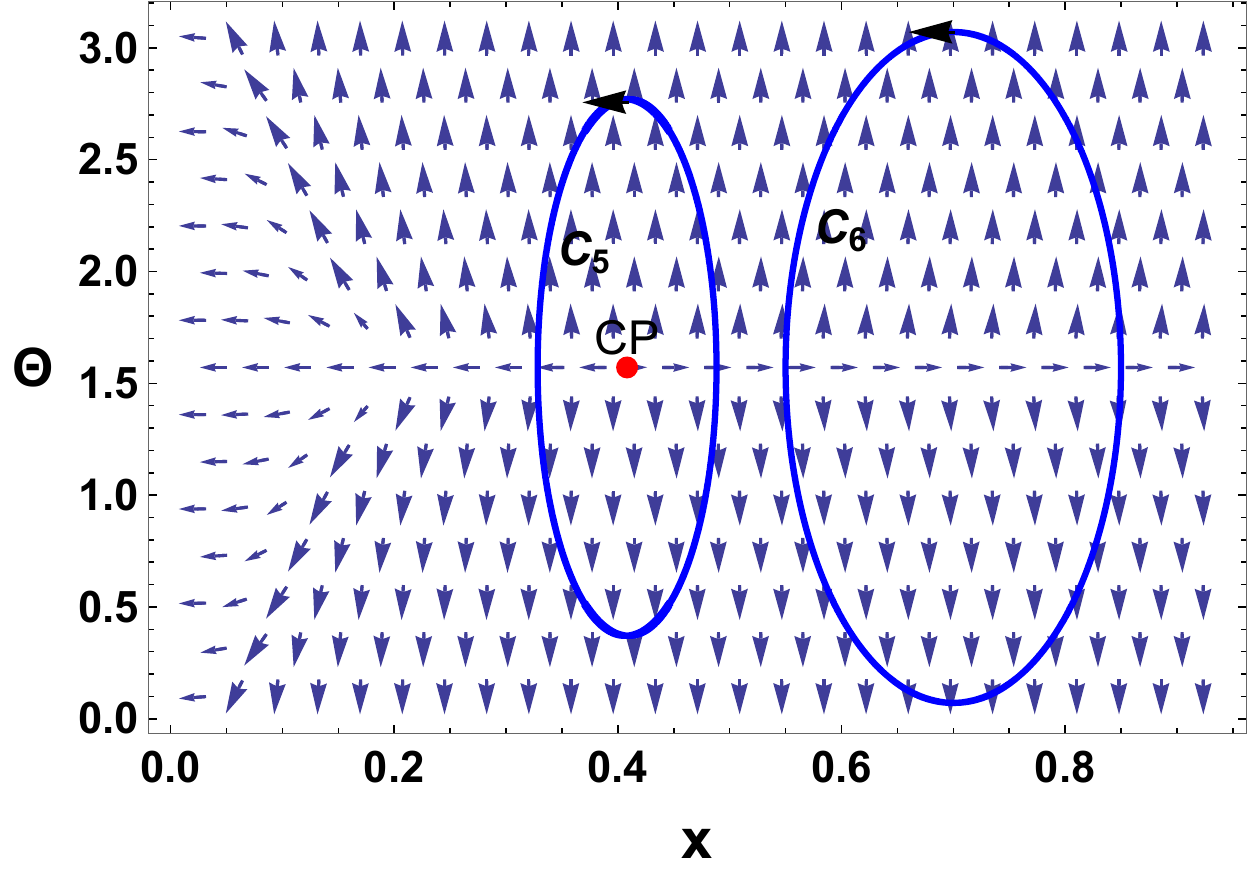}\label{fig:veccneutral}}
    \subfigure[]{\includegraphics[width=0.47 \textwidth]{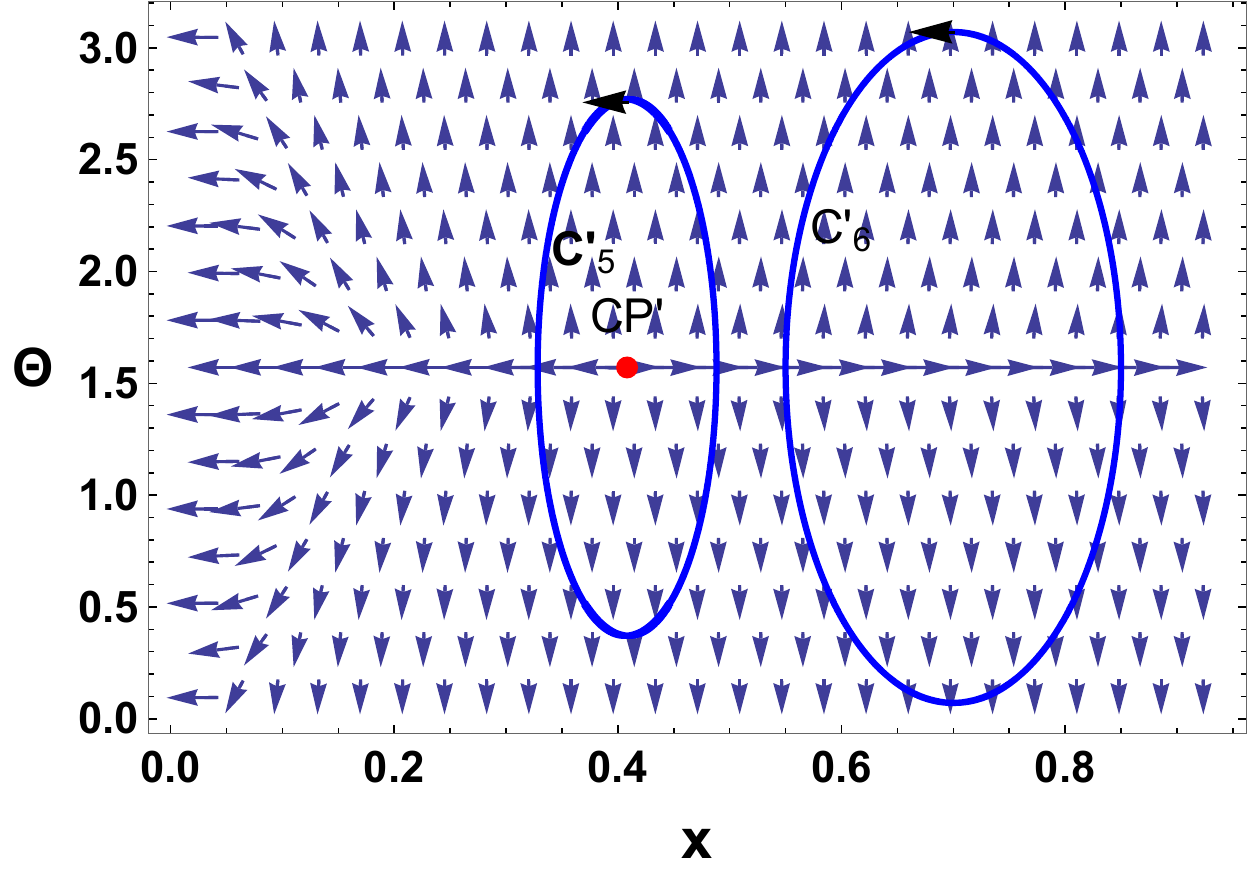}\label{fig:vecccharged}}
    \caption{The arrows depict the unit vector field $\mathbf{n}$ on a portion of the $x-\Theta$ plane, with the zero points indicated by red dots, corresponding to the critical points.  (a) For the dual CFT with $\tilde{Q} = 0$, the zero point of the vector field is located at 
$(x, \Theta) = (1/\sqrt{6}, \pi/2)$.  (b) For the dual CFT with $\tilde{Q} \neq 0$, the zero point of the vector field is located at 
$(x, \Theta) = (0.408257, \pi/2)$ for the charge value $\tilde{Q} = 0.0866855$.  In both cases, we set the boundary radius to $R = 10000$ and the central charge to $A = 27$.}
    \label{veCphase}
\end{figure*}

To study the thermodynamic topology of the critical point, we introduce the thermodynamic potential~\cite{Wei:2021vdx}
\begin{equation}
    \Phi=\frac{\Tilde{T}}{\sin\Theta}=\frac{\csc \Theta \left(24 x^6+4 x^4+y^2\right)}{16 \pi  R x^5}.
\end{equation}
Define the vector field $\phi=(\phi^{\Tilde{S}},\phi^{\Theta})$ as
\begin{equation}
    \phi^{\Tilde{S}}=\left(\frac{\partial \Phi}{\partial \Tilde{S}}\right)_{\Theta,\Tilde{Q},C,A,\mathcal{V}}, \qquad 
    \phi^{\Theta}=\left(\frac{\partial \Phi}{\partial \Theta}\right)_{\Tilde{S},\Tilde{Q},C,A,\mathcal{V}}.
\end{equation}
Then the components of the vector field $\phi$ are
\begin{align}
    \phi^{\Tilde{S}}&=\frac{\csc \Theta \left(24 x^6-4 x^4-5 y^2\right) \left(24 x^6+4 x^4+y^2\right)^2}{32 \pi ^3 L^3 R x^8 \left(24 x^6+12 x^4-y^2\right) \left(4 x^4+3 y^2\right)},\\
    \phi^{\Theta}& =-\frac{\cot \Theta \csc \Theta \left(24 x^6+4 x^4+y^2\right)}{16 \pi  R x^5}.
\end{align}
Evidently, the zero point of the vector field $\phi$ is located at
\begin{align}
\left\{\begin{lgathered}x_{\text{c}}= \frac{1}{6} \left(2^{2/3} K+\frac{2 \sqrt[3]{2}}{K}+2\right)^{\frac{1}{2}},\\
       \Theta=\frac{\pi}{2}.
\end{lgathered}\right.
\end{align}
The results indicate that the zero of the component $\phi^{\Tilde{S}}$ of the vector field coincides with the critical point. In Figs.~\ref{fig:veccneutral} and~\ref{fig:vecccharged},
we present the distribution of the normalized vector field $n=(\phi^{\Tilde{S}}/|| \phi ||, \phi^{\Theta}/|| \phi ||)$ for the thermodynamics of the neutral and charged dual CFTs. 

%================================
%================================

%================================
\begin{figure*}
    \centering
    \subfigure[]{\includegraphics[width=0.47 \textwidth]{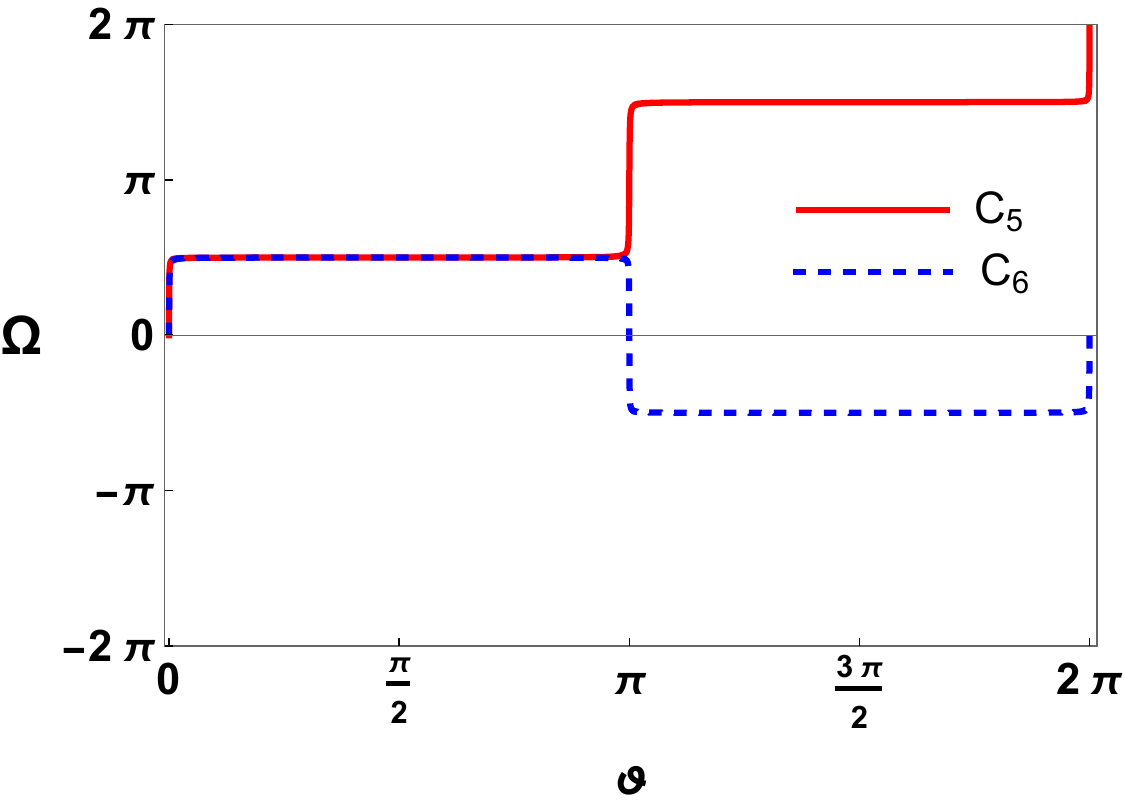}\label{fig:WNCneutral}}
    \subfigure[]{\includegraphics[width=0.47 \textwidth]{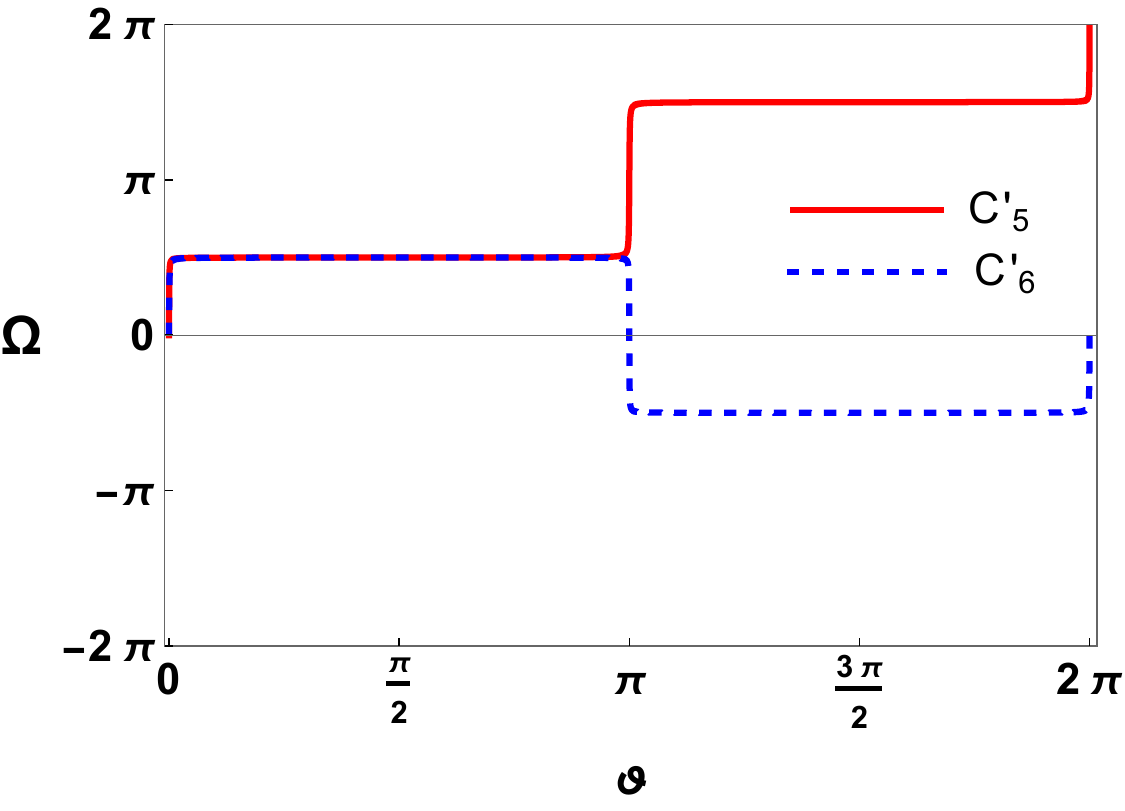}\label{fig:WNCcharged}}
    \caption{$\Omega$ vs.~$\Theta$ for the contours $C_{5}, C_{6}$ in subfigure~(a) 
(with $\tilde{Q} = 0$) and $C'_{5}, C'_{6}$ in subfigure~(b) 
(with $\tilde{Q} \neq 0$).}
    \label{wnCphase}
\end{figure*}
%================================
 
From a topological perspective, it follows that the contour carries a nonzero topological charge if it encloses the critical point and vanishes otherwise. Since the vector field points outward at $\Theta=0$ and $\Theta=\pi$, the total topological charge associated with the thermodynamical critical point is determined by the behavior of the vector field at the lower boundary and at infinity in $x$, which correspond respectively to the lower and upper bounds of the entropy of the dual CFT. To evaluate the topological charge, we consider two contours,  $C_5$ and $C_6$, for the thermodynamics of the neutral dual CFT, and $C'_5$ and $C'_6$  for the charged dual CFT. The corresponding results are displayed in Figs.~\ref{fig:WNCneutral} and~\ref{fig:WNCcharged}. 
Specifically, the topological charges for the contours $C_5$ or $C'_5$, which enclose the critical points $\text{CP}$ and $\text{CP}'$, are found to be $\mathcal{Q}_{\text{CP}}=1$. In contrast, the topological charge associated with $C_6$ and $C'_6$ vanishes, since these contours do not enclose a critical point. Therefore, the total topological charges corresponding to the thermodynamical critical point of both the neutral and charged dual CFT are $\mathcal{Q}_{\text{CP}}=1$. More importantly, the topological charges of the thermodynamics of the dual CFT coincide with those of the bulk thermodynamics~\cite{Jeon:2024yey}.

%================================
\section{Conclusion and discussion}
%================================

In this paper, motivated by the recent proposal that interprets the conformal factor of the AdS boundary metric as a thermodynamic parameter, we established an exact duality between the extended thermodynamics of five-dimensional Gauss–Bonnet black holes and their dual CFT description. For the four-dimensional CFT dual to five-dimensional Gauss–Bonnet gravity with a negative cosmological constant, two independent central charges arise from the trace anomaly. Within our formulation of the CFT first law of thermodynamics, these central charges naturally enter as independent thermodynamic variables, thereby enabling a natural construction of the holographic dictionary.

The thermodynamics of the dual CFT exhibits a phase structure analogous to that of the bulk black hole system. In the fixed $(\Tilde{Q},C,A,\mathcal{V})$ ensemble, the dual CFT displays an oscillatory curve in the $\Tilde{T}-\Tilde{S}$ plane, while its free energy–temperature diagram features the characteristic swallowtail structure below the critical point, signifying a first-order phase transition. Both the oscillatory behavior and the swallowtail structure disappear above the critical point. Furthermore, the phase transition and the critical point in the CFT thermodynamics coincide exactly with those of the five-dimensional charged Gauss–Bonnet AdS black hole.

Finally, we investigated the thermodynamic topology associated with the first-order phase transition and the critical point of the dual CFT. For both neutral and charged dual CFTs, the thermodynamic topology of the first-order phase transition is characterized by a total topological charge $\mathcal{Q}=1$. At the corresponding thermodynamic critical point, the total topological charge is likewise $\mathcal{Q}_{\text{CP}}=1$ in both cases. These topological features of the first-order phase transition and the critical point are identical to those found in the bulk thermodynamics. Our results demonstrate that the thermodynamic behavior of the dual CFT faithfully mirrors that of the five-dimensional charged Gauss–Bonnet AdS black hole, thereby further reinforcing the holographic correspondence between bulk and boundary thermodynamics.

Central charges originate from trace anomalies that arise within the symmetry algebra of a conformal field theory. The anomaly only depends on the dimension of the spacetime on which the conformal field theory resides. For higher-dimensional gravitational theories, multiple central charges may emerge. In five-dimensional Einstein gravity, the dual four-dimensional CFT possesses two identical central charges, whereas in more general gravitational theories, these central charges can take distinct values.
Here, we formulated the holographic thermodynamics of the five-dimensional Gauss–Bonnet gravity, which includes two independent central charges. Extending this construction to higher-dimensional Gauss–Bonnet gravity and exploring the associated holographic first law remain promising avenues for future research.

Thermodynamic topology provides a universal framework for characterizing the global structure of thermodynamic phase space by assigning robust topological invariants to distinct phases. Within the context of AdS black holes and their holographic dual CFTs, this perspective reveals a striking bulk-boundary correspondence: the topological structure encoded in the black hole phase diagram is faithfully mirrored in the thermodynamic behavior of the boundary CFT. Specifically, we have demonstrated that the boundary CFT shares identical topological charges with its bulk counterpart in the case of the five-dimensional charged Gauss-Bonnet AdS black hole. This thermodynamic equivalence further reinforces the holographic connection between gravity and field theory, suggesting that critical phenomena in both the bulk and the boundary belong to the same universality class. Ultimately, these findings deepen our understanding of the geometric and topological foundations of holographic thermodynamics, opening promising avenues for exploring universal topological features across diverse gravitational systems and quantum many-body theories.

Looking forward, it is imperative to extend this framework to higher-order gravitational models, such as quasi-topological and Lovelock gravity. Incorporating additional higher-curvature couplings will naturally yield a more expansive thermodynamic parameter space, giving rise to richer phase structures and novel critical phenomena. Investigating the dual boundary thermodynamics and corresponding topological features within these higher-dimensional, multi-parameter spaces will provide crucial insights into the broader mechanisms of holographic thermodynamics in modified gravity theories.

%===========================================
\acknowledgments

We are grateful to Zi-Qing Chen, Robert B. Mann and Yi Pang for valuable help and useful discussions. We sincerely
thank the anonymous referee for their valuable suggestions, which have
significantly improved the manuscript. This work was supported by the National Natural Science Foundation of China (Grants No. 12305065, No. 12475056,  No. 12475055 and No. 12247101), the China Postdoctoral Science Foundation (Grant No. 2023M731468), the Gansu Province's Top Leading Talent Support Plan, the Fundamental Research Funds for the Central Universities (Grant No. lzujbky-2025-jdzx07), the Natural Science Foundation of Gansu Province (No. 22JR5RA389, No.25JRRA799), and the `111 Center' under Grant No. B20063.
%===========================================

%===========================================
\section*{Appendix}

The exact expressions for the chemical potentials $\mu_C$ and $\mu_A$ are
\begin{subequations}
\label{eq:chemical_potentials}
\begin{align}
    \mu_A &= \frac{\sqrt{\zeta+1}}{\Gamma} \bigg\{ 
    -12 \Big[ 200 \lambda ^4 + (260 \zeta-718) \lambda ^3 + (481-307 \zeta) \lambda ^2 + 6 (16 \zeta-19) \lambda\nonumber \\
    & - 9 \zeta + 9 \Big] x^8 - 12 \Big[ -240 \lambda ^5+ 4 (66 \zeta-401) \lambda ^4 - 22 (47 \zeta-88) \lambda ^3 + (547 \zeta-757) \lambda ^2 \nonumber \\
    &- 6 (19 \zeta-22) \lambda + 9 (\zeta-1) \Big] x^6 + 36 \lambda \Big[ 280 \lambda ^4 + (292 \zeta-722) \lambda ^3 \nonumber \\
    &+ (449-287 \zeta) \lambda ^2 + 18 (5 \zeta-6) \lambda - 9 \zeta + 9 \Big] x^4 \nonumber \\
    &+ x^2 \Big[ 72 \lambda ^2 \big( 40 \lambda ^4 + (76 \zeta-238) \lambda ^3 + (171-109 \zeta) \lambda ^2+ (34 \zeta-40) \lambda - 3 \zeta + 3 \big) \nonumber \\
    &+ \big( -200 \lambda ^4 + (718-260 \zeta) \lambda ^3 + (307 \zeta-481) \lambda ^2+ (114-96 \zeta) \lambda + 9 (\zeta-1) \big) y^2 \Big] \nonumber \\
    &- 6 \lambda \Big[ 120 \lambda ^4 + 2 (54 \zeta-121) \lambda ^3 + (139-89 \zeta) \lambda ^2+ (28 \zeta-34) \lambda - 3 \zeta + 3 \Big] y^2 \bigg\}, \\[10pt]
    \mu_C &= \frac{\zeta^2\sqrt{\zeta+1}}{\Upsilon} \bigg\{ 
    -60 \Big[ 2 \lambda^3 + 3 (\zeta-3) \lambda^2 + (6-4 \zeta) \lambda + \zeta-1 \Big] x^8 + 12 \Big[ 12 \lambda^4 \nonumber \\
    &+ 2 (\zeta+18) \lambda^3 + (39 \zeta-89) \lambda^2 + (40-30 \zeta) \lambda + 5 (\zeta-1) \Big] x^6+ 36 \lambda \Big[ 14 \lambda^3 \nonumber\\
    &+ (19 \zeta-53) \lambda^2 + (32-22 \zeta) \lambda + 5 (\zeta-1) \Big] x^4 + x^2 \Big[ 24 \lambda^2 \big( 6 \lambda^3 + (11 \zeta-37) \lambda^2 \nonumber \\
    &+ (28-18 \zeta) \lambda+ 5 (\zeta-1) \big) - 5 \big( 2 \lambda^3 + 3 (\zeta-3) \lambda^2 + (6-4 \zeta) \lambda + \zeta-1 \big) y^2 \Big] \nonumber \\
    &- 2 \lambda \Big[ 18 \lambda^3 + (23 \zeta-61) \lambda^2 + (34-24 \zeta) \lambda + 5 (\zeta-1) \Big] y^2 \bigg\},
\end{align}
\end{subequations}
where the auxiliary parameters are defined as
\begin{align}
    \zeta &= \sqrt{1-4\lambda}, \\
    \Gamma &= 96 \sqrt{2} \lambda ^3 \left[ 20 \lambda ^2 + (13 \zeta-19) \lambda - 3 \zeta + 3 \right] R x^2 (2 \lambda + x^2), \\
    \Upsilon &= 32 \sqrt{2} \lambda ^3 \left[ 4 \lambda ^2 + (3 \zeta-5) \lambda - \zeta + 1 \right] R x^2 (2 \lambda + x^2).
\end{align}
%===========================================

%===========================================
% \bibliographystyle{JHEP}
% \bibliography{ref}
%===========================================
\providecommand{\href}[2]{#2}\begingroup\raggedright\endgroup
%===========================================
\end{document}